\documentclass[aps,prl,reprint,superscriptaddress]{revtex4-2}

\pdfoutput=1
\usepackage[english]{babel}

\usepackage[utf8]{inputenc}
\usepackage[T1]{fontenc}
\usepackage{lineno}
\usepackage{float}
\usepackage{slashed}
\usepackage{rotating}
\usepackage{amsmath,slashed,xcolor,cancel}
\usepackage{array,booktabs,colortbl,multirow,tablefootnote}
\usepackage{makecell}
\usepackage{chngcntr}
\usepackage{relsize}
\usepackage{colortbl,colordvi,color}
\usepackage{caption}
\usepackage{subcaption}
\usepackage{todonotes}
\usepackage{placeins}
\usepackage{dcolumn}

\usepackage{eurosym}
\usepackage{amssymb}
\usepackage{graphicx}
\usepackage{wrapfig,mdframed,caption}
\usepackage{comment}
\usepackage{overpic} 
\usepackage{xspace} 
\usepackage{ulem} 
\newcommand{\ETmiss}{E_\textrm{T}^\textrm{miss}}

\def\lsim{\mathrel{\rlap{\lower3pt\hbox{\hskip0pt$\sim$}}
   \raise1pt\hbox{$<$}}}         
\def\gsim{\mathrel{\rlap{\lower4pt\hbox{\hskip1pt$\sim$}}
   \raise1pt\hbox{$>$}}}         

\def\cw{c_{\tilde{W}}}
\def\cb{c_{\tilde{B}}}
\def\cg{c_{\tilde{G}}}

\def\pt{p_\textrm{T}}
\def\mll{M(\ell\ell)}
\newcommand*{\GeV}{\ensuremath{\text{Ge\kern -0.1em V}}}
\newcommand*{\TeV}{\ensuremath{\text{Te\kern -0.1em V}}}

\newcommand{\commentsinout}[1]{}
\newcommand{\PQs}[1]{\commentsinout{\textbf{{\color{teal} #1}}}} 

\begin{document}

\title{Constraining off-shell production of axion-like particles with $Z\gamma$ and $WW$ differential cross-section measurements}

\author{Sonia Carr{\'a}}
\author{Vincent Goumarre}
\author{Ruchi Gupta}
\author{Sarah Heim}
\affiliation{Deutsches Elektronen-Synchrotron DESY, 22607 Hamburg, Germany}
\author{Beate Heinemann}
\affiliation{Deutsches Elektronen-Synchrotron DESY, 22607 Hamburg, Germany}
\affiliation{Physikalisches Institut, Albert-Ludwigs-Univerit{\"at} Freiburg, 79104 Freiburg, Germany}
\author{Jan K{\"u}chler}
\author{Federico Meloni}
\author{Pablo Quilez}
\author{Yee-Chinn Yap}
\affiliation{Deutsches Elektronen-Synchrotron DESY, 22607 Hamburg, Germany}

\date{\today}

\begin{abstract}
This article describes a search for low-mass axion-like particles (ALPs) at the Large Hadron Collider (LHC).
If ALPs were produced at the LHC via gluon-gluon fusion and decayed to bosons, the energy dependence of the measured diboson cross-sections would differ from the Standard Model expectation. Measurements of $WW$ and $Z\gamma$ differential cross-sections by the ATLAS collaboration are interpreted to constrain ALP 
couplings to $W$-, $Z$-bosons and photons assuming gluon-gluon-fusion production. 
\end{abstract}

\maketitle

Axions and more generally axion-like particles (ALPs)~\cite{Georgi:1986df,Choi:1986zw} appear in many extensions of the Standard Model (SM). Often introduced to solve specific questions like the strong CP problem~\cite{Peccei:1977hh,Peccei:1977ur,Weinberg:1977ma,Wilczek:1977pj}, they can also be promising dark matter candidates~\cite{Preskill:1982cy, Abbott:1982af,Dine:1982ah}. The canonical QCD axion is expected to be extremely light, $m_a\lesssim 10^{-2}\,\text{eV}$, and feebly interacting, with axion scales out of direct experimental reach, $f_a\gtrsim 10^8 \,\text{GeV}$. However, recently proposed ``heavy QCD axions'' motivated by the Peccei-Quinn (PQ) quality problem~\cite{Holman:1992us,Kamionkowski:1992mf, Barr:1992qq, Ghigna:1992iv, Georgi:1981pu, Giddings:1988cx,Coleman:1988tj,Gilbert:1989nq, Rey:1989mg} may involve scales as low as $f_a\sim$TeV~\cite{rubakov:1997vp,Berezhiani:2000gh,Gianfagna:2004je,Hsu:2004mf,Hook:2014cda,Fukuda:2015ana,Chiang:2016eav,Dimopoulos:2016lvn,Gherghetta:2016fhp,Kobakhidze:2016rwh,Agrawal:2017ksf,Agrawal:2017evu,Gaillard:2018xgk,Buen-Abad:2019uoc,Csaki:2019vte,Gupta:2020vxb,Gherghetta:2020ofz,Takahashi:2021tff}, renewing the interest of collider searches for the axion~\cite{Jaeckel:2015jla,Mimasu:2014nea,Bauer:2017ris,Bauer:2018uxu,Brivio:2017ije,Craig:2018kne,CidVidal:2018blh,Izaguirre:2016dfi,Freytsis:2009ct,Alonso-Alvarez:2018irt,Gavela:2019cmq,Hook:2019qoh,Haghighat:2020nuh,Alonso-Alvarez:2021ett,Chakraborty:2021wda}.

Searches for ALPs have been performed at many different experiments and cover a large range of masses ($m_a$) and coupling strengths to SM particles; see Ref.~\cite{Strategy:2019vxc} for a recent compilation of results. Following the suggestion of Ref.~\cite{Gavela:2019cmq}, in this article we explore instead the high-energy tails of differential diboson cross-sections in the search for off-shell production of low-mass ALPs. If $\sqrt{\hat{s}}$ is the energy of the hard-scatter interaction, the cross-section for $s$-channel production of boson pairs via ALP exchange is expected to increase with $\hat{s}$, while the SM cross-sections fall with $1/\hat{s}$.

The theoretical framework used throughout this article is a linear effective field theory (EFT), in which  electroweak physics beyond the SM (BSM) is described by a linear EFT  expansion~\cite{Buchmuller:1985jz,Grzadkowski:2010es} in terms of towers of gauge invariant operators ordered by their mass dimension. 
The chosen EFT includes the SM plus an ALP~\cite{Georgi:1986df,Choi:1986zw,Brivio:2017ije}, and the scale of the new physics is the ALP decay constant $f_a$. The EFT approach is only valid if the probed energy is much lower than this scale, i.e. $\sqrt{\hat{s}}\ll f_a$. However, in the applied model the value of $f_a$ only affects the overall cross-section and not the differential distribution, so that the results can easily be scaled to any value of $f_a$. 

The most general CP-conserving effective Lagrangian describing bosonic ALP couplings reads  
\begin{multline}
    \mathcal{L}=\frac{1}{2}\partial_{\mu} a\partial^{\mu} a  + \frac{1}{2}m_a^2a^2 +\frac{1}{4} g_{a g g} a G \tilde{G}+\frac{1}{4} g_{a W W} a W \tilde{W}\\
    +\frac{1}{4} g_{a Z Z} a Z \tilde{Z}     +\frac{1}{4} g_{a \gamma \gamma} a F \tilde{F}+\frac{1}{4} g_{a \gamma Z} a F \tilde{Z}\,.
\end{multline}
These bosonic interactions depend solely on three coefficients $\cw$, $\cb$ and $\cg$, which can be directly related to physical interactions and to the coupling parameters~\cite{Gavela:2019cmq}:

\begin{eqnarray}
\label{eq:couplings}
g_{agg} &= &\frac{4\cg}{f_a} \mbox{ , } g_{aWW}=\frac{4\cw}{f_a} \\
g_{a\gamma\gamma} &= &\frac{4}{f_a}\left(\sin^2\theta_W\cw+\cos^2\theta_W\cb\right)\\
g_{aZZ} &= &\frac{4}{f_a}\left(\sin^2\theta_W\cb+\cos^2\theta_W\cw\right)\\
g_{aZ\gamma} &= &\frac{8}{f_a}\sin\theta_W\cos\theta_W(\cw-\cb),
\end{eqnarray}
where $g_{agg}$ is the coupling strength of the ALP to gluons, $g_{a\gamma\gamma}$ to photons, $g_{aWW}$/$g_{aZZ}$  to $W$/$Z$ bosons and $g_{aZ\gamma}$ to a $Z$ boson and a photon.   
An additional term that couples the ALP to the Higgs field can be introduced in the bosonic Lagrangian inducing a mixing of the ALP with the longitudinal component of the $Z$ boson, and affecting the fermion couplings. For this article the coefficient of this term, $c_{a\phi}$ in Refs.~\cite{Gavela:2019wzg,Gavela:2019cmq}, is set to zero~\footnote{It was also tested that for $c_{a\phi}=1$ the same results are obtained.}. 

In Ref.~\cite{Gavela:2019cmq}, diphoton, dijet and $ZZ$ data were used to constrain the parameters $g_{agg}$, $g_{a\gamma\gamma}$ and $g_{aZZ}$. 
In this article, the differential cross-sections for $WW$ and $Z\gamma$ production measured with the ATLAS detector~\cite{Aaboud:2019nkz,Aad:2021dse,Aad:2019gpq} are analyzed to constrain $g_{aWW}$ and $g_{aZ\gamma }$. The ALPs are assumed to be produced via the gluon-gluon fusion (ggF) process. In principle other production modes, such as vector-boson fusion, are also possible but are not considered in this article.

In the regime $\hat{s}\gg m_a^2$ and $\hat{s}\gg m_V^2$, the ggF cross-section for non-resonant production of two bosons, $V_1$ and $V_2$, mediated by an ALP, is given by 
\begin{equation}
    \sigma(V_1V_2)\propto g_{agg}^2g_{aV_1V_2}^2 \hat{s} 
\end{equation}
where $g_{agg}$ is the coupling of the ALP to gluons and $g_{aV_1V_2}$ is the coupling to the two vector bosons of the ALP decay. 

Three measurements are interpreted in this article: a $WW$ cross-section measurement with a jet veto ($WW0j$)~\cite{Aaboud:2019nkz}, a $WW$ cross-section measurement with a requirement of at least one jet ($WW1j$)~\cite{Aad:2021dse} and a $Z\gamma$ cross-section measurement~\cite{Aad:2019gpq}. Only events in which one $W$ boson decays to an electron and an electron neutrino ($e\nu_e$) and the other to a muon and a muon neutrino ($\mu\nu_\mu$) are considered in the $WW$ measurements, while the $Z\gamma$ analysis only includes $Z$ boson decays to electron or muon pairs. Corrections are applied for any experimental effects and the results are presented in fiducial regions defined by the selection criteria in Table~\ref{tab:selcuts}.

The $WW0j$ measurement uses data with an integrated luminosity of $36.1$~fb$^{-1}$. Two leptons are selected with transverse momentum $\pt(\ell)>27$~GeV and a requirement on the pseudorapidity of $|\eta(\ell)|<2.5$. Requirements on the invariant mass of the two leptons, $\mll$, and the transverse momentum of the dilepton system, $\pt(\ell\ell)$, are applied to reduce background due to Drell-Yan production, as well as the contribution of Higgs boson decays. The magnitude of the missing transverse momentum, $\vec{p}_\textrm{T}^{miss}$, is denoted as $\ETmiss$. Events with hadronic jets with $\pt>35$~GeV and $|\eta|<4.5$ are vetoed to suppress backgrounds from top quark production. The measurement is performed differentially in six observables. 

The $WW1j$ measurement is based on $139$~fb$^{-1}$ of data. In addition to kinematic requirements on the leptons and the dilepton system, events are required to have at least one jet with $\pt>30$~\GeV\ and $|\eta|<4.5$. The measurement is performed differentially in numerous variables. 

\begin{table}[htbp]
    \centering
    \caption{ \label{tab:selcuts} Selection criteria defining the fiducial regions of the $WW0j$, $WW1j$ and $Z\gamma$ 
    cross-section measurements~\cite{Aad:2019gpq,Aad:2021dse, Aaboud:2019nkz}. In the $Z\gamma$ analysis the photon 
    must be isolated w.r.t. hadrons in the event, see text and Ref.~\cite{Aaboud:2019nkz}. Only jets with $|\eta(\textrm{jet})|<4.5$ are considered for the jet $\pt$ requirements.}
    \begin{tabular}{l  c   c   c}
    \hline\hline
    Variable & \multicolumn{3}{c}{Selection Cut} \\
     & $WW0j$  & $WW1j$  & $Z\gamma$  \\\hline
         $\pt(\ell)$ [\GeV]& $>27$ & $>27$ &$>30,25$ \\
         $|\eta(\ell)|$ & $<2.5$ & $<2.5$ & $<2.47$\\
         $\mll$ [\GeV]& $>55$ & $>85$ & $>40$\\
         $\pt(\ell\ell)$ [\GeV] & $>30$ & -- & --\\
         $\ETmiss$ [\GeV]& $>20$ & -- & --\\
         $\pt(\textrm{jet})$ [GeV] & $<35$ & $>30$ & --\\
         $\pt(\gamma)$ [\GeV]& -- & -- & $>30$ \\
         $|\eta(\gamma)|$ & -- & -- & $<2.37$ \\
         $\Delta R(\ell,\gamma)$ & -- & --& $>0.4$ \\
         $M(\ell\ell\gamma)+M(\ell\ell)$ [\GeV]& -- & -- & $>182$ \\
         \hline
    \end{tabular}
  
\end{table}

The $Z\gamma$ cross-section measurement is also based on 139~fb$^{-1}$ of data, and selects events with a $Z$ boson candidate as well as an isolated photon with high transverse momentum, $\pt(\gamma)>30$~GeV and a requirement on the photon pseudorapidity of $|\eta(\gamma)|<2.37$. A requirement on the sum of the invariant masses of the dilepton system and the $\ell\ell\gamma$ system, $M(\ell\ell\gamma)$,  is applied to reduce the contribution from events where the photon is radiated off a lepton. The photon is required to be isolated from hadrons and leptons within a cone defined by $\Delta R=\sqrt{\Delta\eta^2+\Delta\phi^2}$ where $\Delta \eta$ and $\Delta\phi$ are the differences between the photon and the other particle in pseudo-rapidity and azimuthal angle, respectively. The $Z\gamma$ cross-sections are measured differentially in six observables. 

Cross-sections for SM $q\bar{q}\to WW$ production are determined at next-to-next-to-leading order (NNLO) in QCD, using the parton-level generator \textsc{Matrix}~\cite{Gehrmann_2015,Cascioli_2012,Ball:2017nwa}, including off-shell effects and the non-resonant and resonant gluon-initiated contributions at LO. For improved precision, the NNLO MATRIX prediction is also complemented with NLO corrections to gluon-induced WW production~\cite{Caola:2016trd} and with NLO electroweak (EW) corrections that also include the photon-induced contribution~\cite{Biedermann:2016guo}. For all predictions the NNPDF 3.1 LUXqed parton distribution function (PDF) set is used~\cite{Ball:2017nwa,Bertone:2017bme}. The renormalization and factorization scales are set to half the invariant mass of the $W$ bosons, $M(WW)/2$. 

The SM cross-sections for the $\ell \ell \gamma$ process are computed with \textsc{Matrix}~\cite{Grazzini_2018} at NNLO in QCD, corrected to particle level using the factors provided in Ref.~\cite{Aad:2019gpq}. The gluon-initiated contributions is included at LO but amounts to only 2\%. 
The cross-sections are obtained with the CT14nnlo PDF set~\cite{Dulat_2016}, and the transverse momentum ($q_T$) subtraction
method~\cite{PhysRevLett.98.222002}. The values of the renormalisation and factorization scales are set to $\sqrt{\mll)^2+\pt(\gamma)^2}$. Smooth-cone photon isolation~\cite{Frixione_1998} is imposed as discussed in Ref.~\cite{Aad:2019gpq}. EW radiative corrections to $Z\gamma$ production have been computed at NLO
for the $q\bar{q}$, $q\gamma$ and $\gamma\gamma$ initial states in Refs.~\cite{Denner_2016,Hollik_2004,Accomando_2006}. 

The absence of a complete combined calculation of NLO EW and NNLO QCD corrections leads to the question whether the NLO EW corrections associated with the $q\bar{q}$ initial state should be applied multiplicatively or additively to the NNLO QCD corrections. We follow the strategy of the respective publications: For $WW0j$ and $Z\gamma$, the average between the two choices is taken as central value and the difference between the average and either of the choices as the uncertainty, as recommended in Ref.~\cite{Kallweit:2019zez}. In the $WW1j$ analysis, the NLO EW corrections are applied multiplicatively and no uncertainty is applied as in the probed phase space the electroweak and QCD corrections should factorize. 

Uncertainties on the theoretical predictions due to possible higher order QCD contributions are estimated by varying the renormalisation and factorization scales by factors of up to two. Uncertainties arising from the choice of PDF set and the value of $\alpha_s$ are assessed according to the PDF4LHC recommendations~\cite{Butterworth_2016} using the 68\% confidence level (CL) variations of the NNPDF set~\cite{Ball_2014}.

ALP signal events are produced with the \textsc{MadGraph5}\_aMC@NLO~\cite{Alwall:2014hca} Monte Carlo generator interfaced to \hbox{\textsc{Pythia}~8~\cite{Sjostrand:2014zea}} for parton showering and hadronization. Samples are produced varying either $\cg,~\cb$ or $\cw$ for a fixed value of $f_{a}$ using the set of operators described in Ref.~\cite{Brivio:2017ije}. The following parameters are used unless otherwise stated: $f_a=1$~\TeV and $m_a=1$~keV. 
For this analysis the value of $m_a$ is irrelevant as long as $m_a\ll \hat{s}$, and it was explicitly tested that the cross-section is independent of $m_a$ for $m_a<100$~\GeV. The coupling of the axion to gluons, $g_{agg}$, is in principle an independent free parameter. In the simulation it was set to $g_{agg}=1$~TeV$^{-1}$ in line with Ref.~\cite{Gavela:2019cmq}, and some of the results below are presented for this assumption.
However, in order to compare the sensitivity of this analysis with constraints from previous experiments, it is assumed that the ratio of the EW axion couplings over the gluonic coupling is controlled by the strength of the corresponding gauge coupling constants,  $g_{aV_1V_2}/g_{agg}=\alpha_{V_1V_2}/\alpha_s$ \footnote{$\alpha_{V_1V_2}$ is defined as $\alpha_{e m}=\alpha_{WW} s_{w}^{2}=\alpha_{Z\gamma} s_{w}c_{w}$, where  $s_w$ and $c_w$ denote the sine and cosine of the weak mixing angle}.
This assumption is well motivated by pseudo Nambu-Goldstone bosons with anomalous couplings generated by
the triangle diagram with $\mathcal{O}(1)$ group theory factors, see e.g. 
Ref.~\cite{Alonso-Alvarez:2018irt}.

In the $\ell\ell\gamma$ final state, only $a \rightarrow Z\gamma$ is considered. In principle there is also a contribution from 
$\gamma^*\gamma$  but based on the constraints already set in Ref.~\cite{Brivio:2017ije} it is negligible. The A14 tune is used for the modeling of parton showering, the hadronization and the underlying event. Parton luminosities are provided by the NNPDF2.3LO PDF set. The LO cross-section from the generator prediction is used. It is worth noting that it is likely that there are large NLO corrections, similar to the factor of $\sim$2 found for ggF production of the Higgs boson, and accounting for these would result in more stringent constraints than those presented here. For the ALP predictions no theoretical uncertainties are considered.

The fiducial selections of the $WW$ and $Z\gamma$ cross-section measurements shown in Table~\ref{tab:selcuts} are applied to the generated ALP events using RIVET analysis routines~\cite{Bierlich_2020}. Interference between the SM and the ALP process is neglected.
For the $WW0j$ channel, 17\% of all $a\to WW \to e\nu \mu \nu$ events pass the fiducial selection. Most of the inefficiency arises from the jet veto which rejects nearly 75\% of the ALP events. For the $WW1j$ channel, the fraction of $a\to WW \to e\nu \mu \nu$ events passing the fiducial selection is 45\%.  
In the $Z\gamma$ channel, 55\% of all $a\to Z\gamma \to \ell^+\ell^-\gamma$ ($\ell=e$ and $\mu$) events are selected. 

The measured cross-section in each bin is compared to the sum of the SM and varying ALP contributions. The predictions and the measured cross-sections are used to construct a binned likelihood function ${\cal L}$ based on a product of Poisson probability terms over all bins of the differential cross-section distribution.
This function depends on the product of coupling parameters, $g_{aV_1V_2}\times g_{agg}$, and a set of nuisance parameters $\theta$ that encode the effect of systematic uncertainties in the signal and background expectations. All experimental uncertainties on the measurements as documented in Ref.~\cite{Aaboud:2019nkz,Aad:2019gpq,Aad:2021dse} and the uncertainties on the SM predictions related to PDFs and higher order QCD/EW corrections as discussed above, are incorporated as nuisance parameters with Gaussian constraints. 

The sensitivity to $g_{aV_1V_2}\times g_{agg}$ is studied for each of the measured observables using pseudo-experiments. It is evaluated by determining the expected upper limits on the relevant ALP coupling parameters. 

\begin{figure}
 \includegraphics[width=0.9\columnwidth]{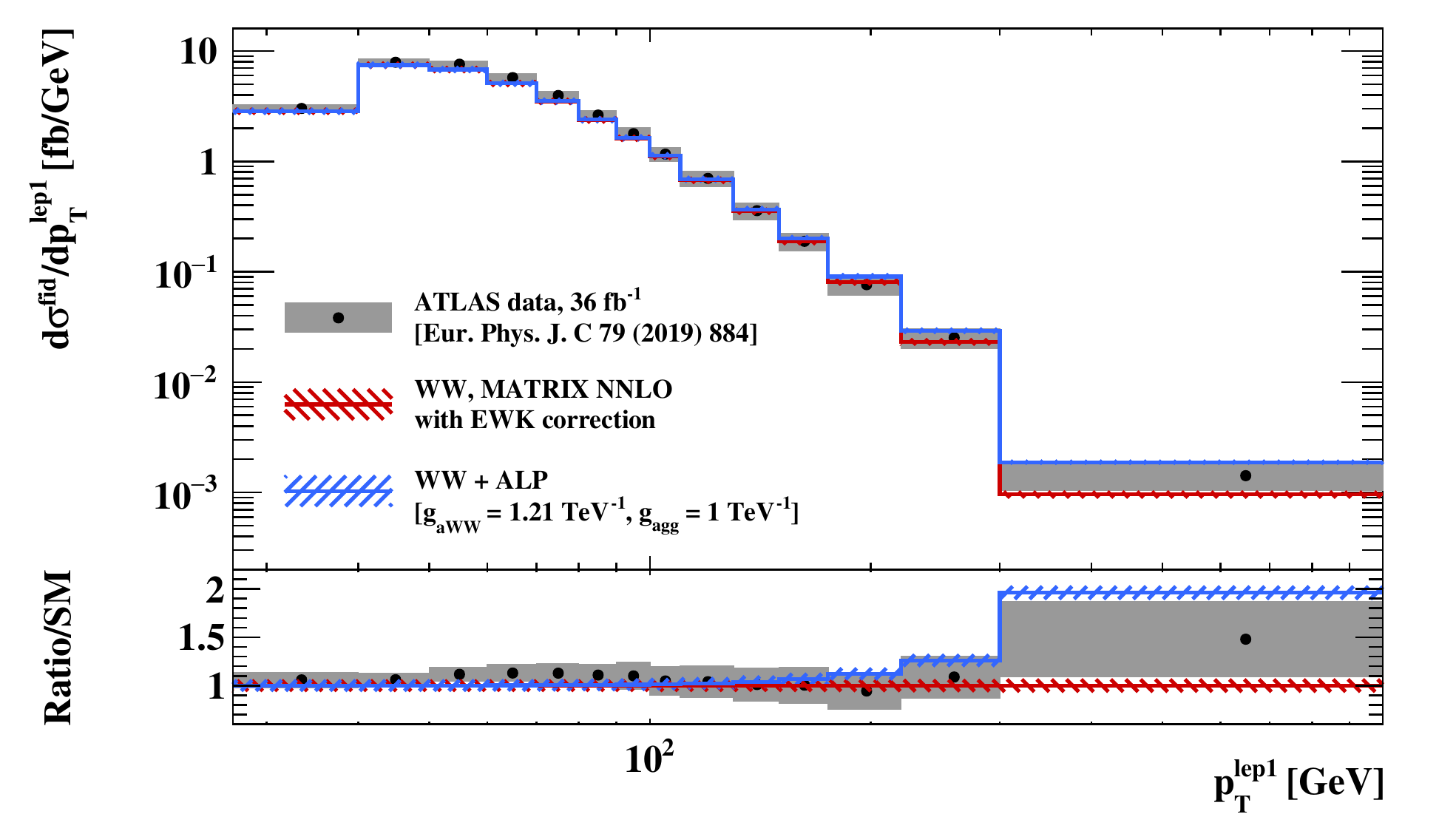}
 \includegraphics[width=0.9\columnwidth]{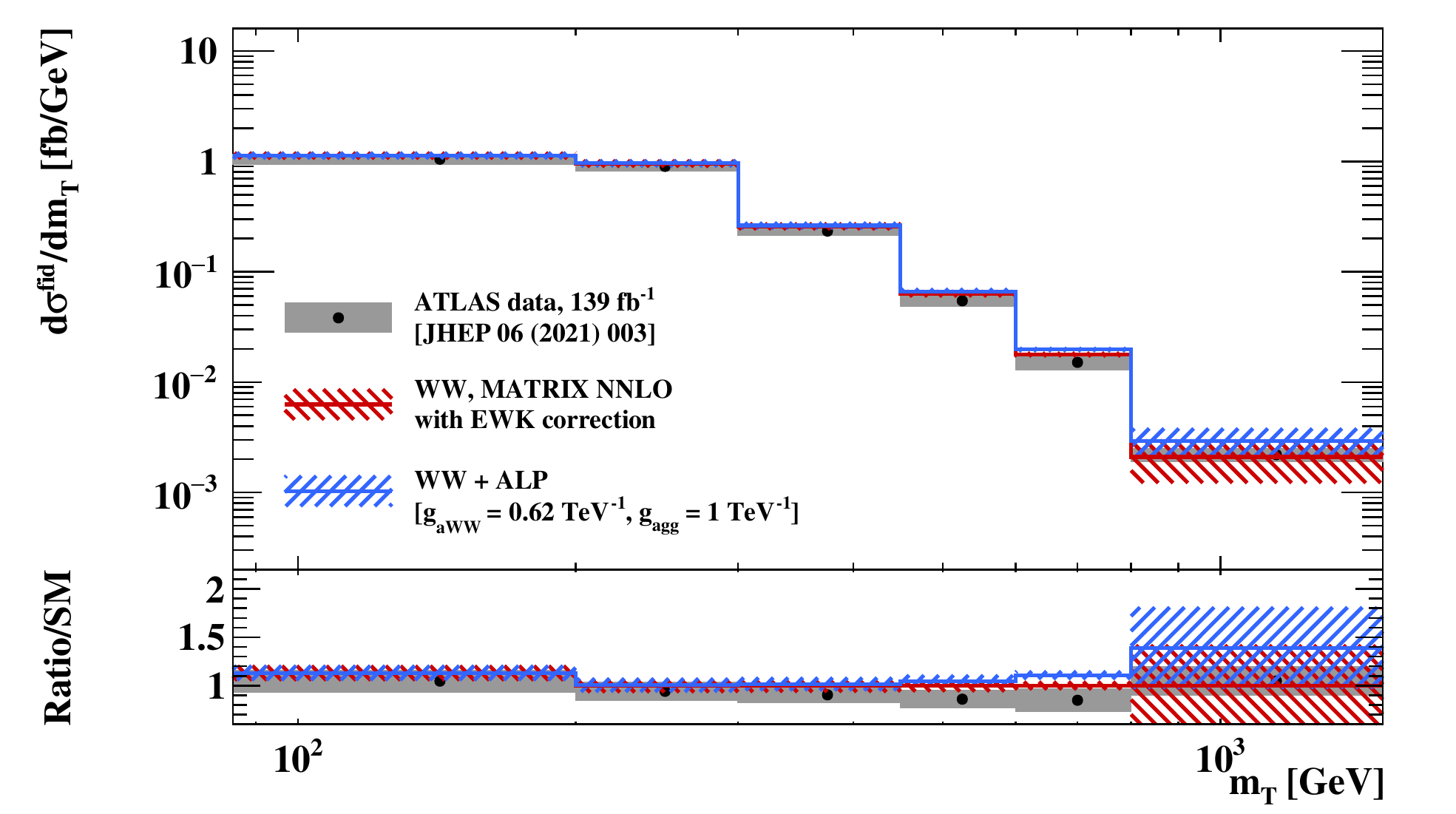}
    \caption{\label{fig:wwplot}Differential cross-sections as a function of (top) the leading lepton $\pt$ measured in $WW0j$ events, (bottom) the transverse mass $m_T$ measured in $WW1j$ events. The data is compared to the SM prediction and a hypothetical ALP signal with $m_a=1$~keV, $f_a=1$~\TeV, $g_{agg} =1$~\TeV$^{-1}$, and $g_{aWW}$ as shown in the legend corresponding to the derived 95\% CL upper limit in each analysis. The lower panels show the ratios of the data and of the sum of SM and ALP contribution to the SM prediction.
    }
\end{figure}

\begin{figure}
     \includegraphics[width=0.9\columnwidth]{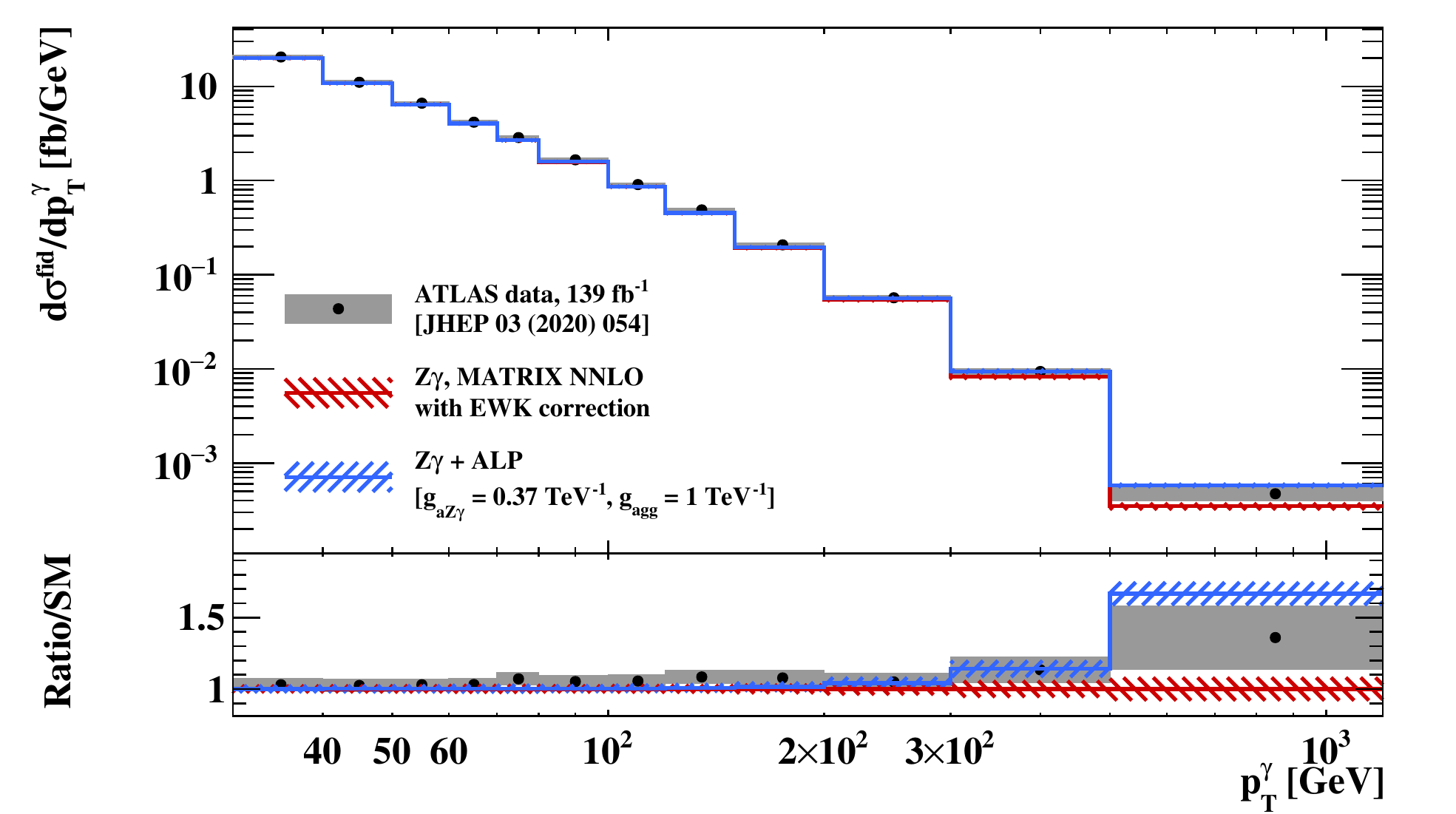}
    \caption{\label{fig:zgplot} Differential cross-sections as a function of 
    the photon $\pt$ measured in $Z\gamma$ events. The data is compared to the SM prediction and a hypothetical ALP signal with $m_a=1$~keV, $f_a=1$~\TeV, $g_{agg} =1$~\TeV$^{-1}$, and $g_{aWW}$ and $g_{aZ\gamma}$ as shown in the legend corresponding to the derived 95\% CL upper limit of the analysis.
    The lower panels show the ratios of the data and of the sum of SM and ALP contribution to the SM prediction.
    }
\end{figure}

For $WW0j$ production, the $\pt$ of the leading lepton provides the best sensitivity, while for $WW1j$ production the transverse mass of the dilepton system and $\ETmiss$, $m_T=\sqrt{(p_\textrm{T}(\ell\ell)+\ETmiss)^2-(\vec{p}_\textrm{T}(\ell\ell)+\vec{p}_\textrm{T}^\textrm{ miss})^2}$ is the most sensitive observable. For $Z\gamma$, the photon $\pt$ is the strongest discriminator. The distributions are shown in Fig.~\ref{fig:wwplot} and Fig.~\ref{fig:zgplot}, compared to the SM prediction and to the sum of the prediction for the SM and a hypothetical ALP signal, for the $WW$ and $Z\gamma$ analyses, respectively. 

The likelihood fits to the data show no significant deviation from the SM predictions: The central values for both the $WW$ and the $Z\gamma$ coupling of the ALP are consistent with $0$ within $<1.5\sigma$. 
The observed and expected limits on the coupling parameters at the 95\% CL are given in Table~\ref{tab:results}. 
The expected upper limit of the $WW1j$ analysis is about a factor two better than that of the $WW0j$ analysis. 

While the data constrain the product, $g_{agg}g_{aVV}$, in Table~\ref{tab:results} constraints on $g_{aVV}$ are presented for a fixed value of $g_{agg}=1$~TeV$^{-1}$. They can trivially be converted into limits on the product of the couplings $g_{agg} g_{aWW}$ and $g_{agg}g_{aZ\gamma}$: the observed upper limits are $g_{aWW}<0.62$~TeV$^{-2}$ and $g_{aZ\gamma}<0.37$~TeV$^{-2}$ at 95\% CL, respectively.

\begin{table}[htbp]
    \centering
        \caption{\label{tab:results} Observed (obs.) and expected (exp.) 95\% CL upper limits on the coupling parameters $g_{aWW}$ and $g_{aZ\gamma}$ based on the fit to the $WW0j$, $WW1j$ and $Z\gamma$ data assuming $g_{agg}=1$~TeV$^{-1}$, respectively. 
        Also given are the corresponding constraints on $\cw$  and $\cw-\cb$ for $f_a=1$~TeV. 
    }
    \begin{tabular}{l   l   c   d  d}
    \hline\hline
    Parameter    & Analysis & $\int {\cal L}\textrm{d}t$ & \multicolumn{2}{c}{95\% CL  upper limit} \\
                  & & [fb$^{-1}$] & \mbox{obs.} & \mbox{exp.} \\\hline
                  
$g_{aWW}$ [TeV$^{-1}$] & $WW0j$, $\pt^{\ell 1}$ & 36 &  1.21 & 1.00 \\
$g_{aWW}$ [TeV$^{-1}$] & $WW1j$, $m_T$ & 139 & 0.62 & 0.49 \\
$g_{aZ\gamma}$ [TeV$^{-1}$] & $Z\gamma$, $\pt^\gamma$ & 139 & 0.37 & 0.29 \\
$|\cw|$     & $WW0j$, $\pt^{\ell 1}$ & 36  &0.30 & 0.25 \\
$|\cw|$     & $WW1j$, $m_T$ & 139 & 0.15 & 0.12 \\
$|\cw-\cb|$ & $Z\gamma$ $\pt^\gamma$ & 139  & 0.11 & 0.09 \\\hline
    \end{tabular}
\end{table}

Based on Eq.~\ref{eq:couplings}, and assuming a value for $f_a$, the limits on the coupling parameters can be translated to constraints on the coefficients $\cw$ and $\cb$, as shown in Table~\ref{tab:results}. The value of $f_a=1$~TeV is chosen so that the results can be compared directly to those derived in Ref.~\cite{Gavela:2019cmq} from $ZZ$ and $\gamma\gamma$ measurements. It is worth noting that the choice of $f_a=1$~TeV is rather controversial as $f_a$ is the scale of new physics and the data used here actually probe that regime directly, making the usage of the EFT questionable. Since the choice of $f_a$ only affects the coupling strength and does not affect the kinematic distributions, the $\cw$ and $\cb$ values can easily be rescaled to any $f_a$ value. For instance, for $f_a=5$~TeV the values of the upper limits on the coefficients are five times higher. 

It was also tested how the expected upper limits of the $WW0j$ and $WW1j$ compare when using the same method and the same luminosity. Using the $\pt^{\ell 1}$ distribution and an integrated luminosity of 139~fb$^{-1}$ for both gives an expected limit of $|\cw|<0.20$ for the $WW0j$ and $|\cw|<0.16$ for the $WW1j$ analysis.
While the $WW1j$ provides a slightly stronger constraint, a statistical combination of the two results would likely yield improved results and could be considered in the future. The expected limit for the $WW1j$ analysis of $|\cw|<0.16$ based on $\pt^{\ell 1}$ is a factor $1.25$ larger than that obtained based on $m_\textrm{T}$, shown in Table~\ref{tab:results}. Thus $m_\textrm{T}$ was used for the data interpretation.

\begin{figure}[htbp]
    \includegraphics[width=\columnwidth]{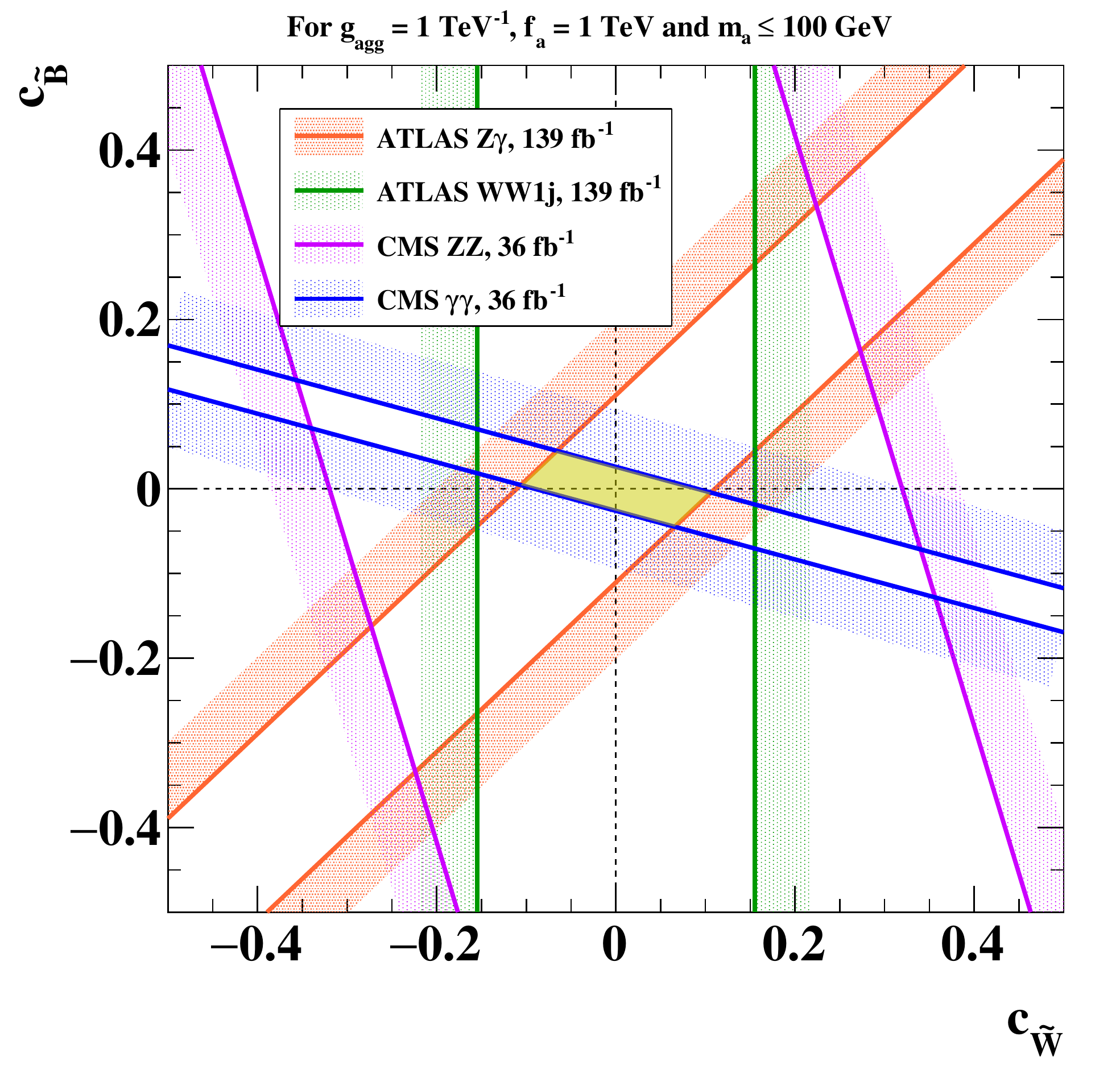}
    \caption{95\% CL constraints on $\cw$ and $\cb$ from this work ($Z\gamma$ and $WW$) and from Ref.~\cite{Gavela:2019cmq} ($ZZ$ and $\gamma\gamma$). Values of $g_{agg}=1$~TeV$^{-1}$, $f_a=1$~TeV and $m_a=1$~keV are assumed. The yellow area is still allowed by all analyses. }
    \label{fig:cbwlimits}
\end{figure}

The observed upper limits on the coefficients $\cb$ and $\cw$ for $f_a=1$~TeV are displayed in Fig.~\ref{fig:cbwlimits} for four analyses: $\gamma\gamma$ and $ZZ$ from Ref.~\cite{Gavela:2019cmq}, and $WW1j$ and $Z\gamma$ from the present analysis. 
The $WW$ process is independent of $\cb$ while the other three processes depend on both $\cw$ and $\cb$ with different dependencies. When considering all four constraints only the small area near zero is allowed with $|\cw|<0.11$ and $|\cb|<0.045$. This area is constrained just by the $\gamma\gamma$ and $Z\gamma$ processes, and the $WW$ and $ZZ$ processes add no further information in this model. 

\begin{figure}[htbp]
    \includegraphics[width=\columnwidth]{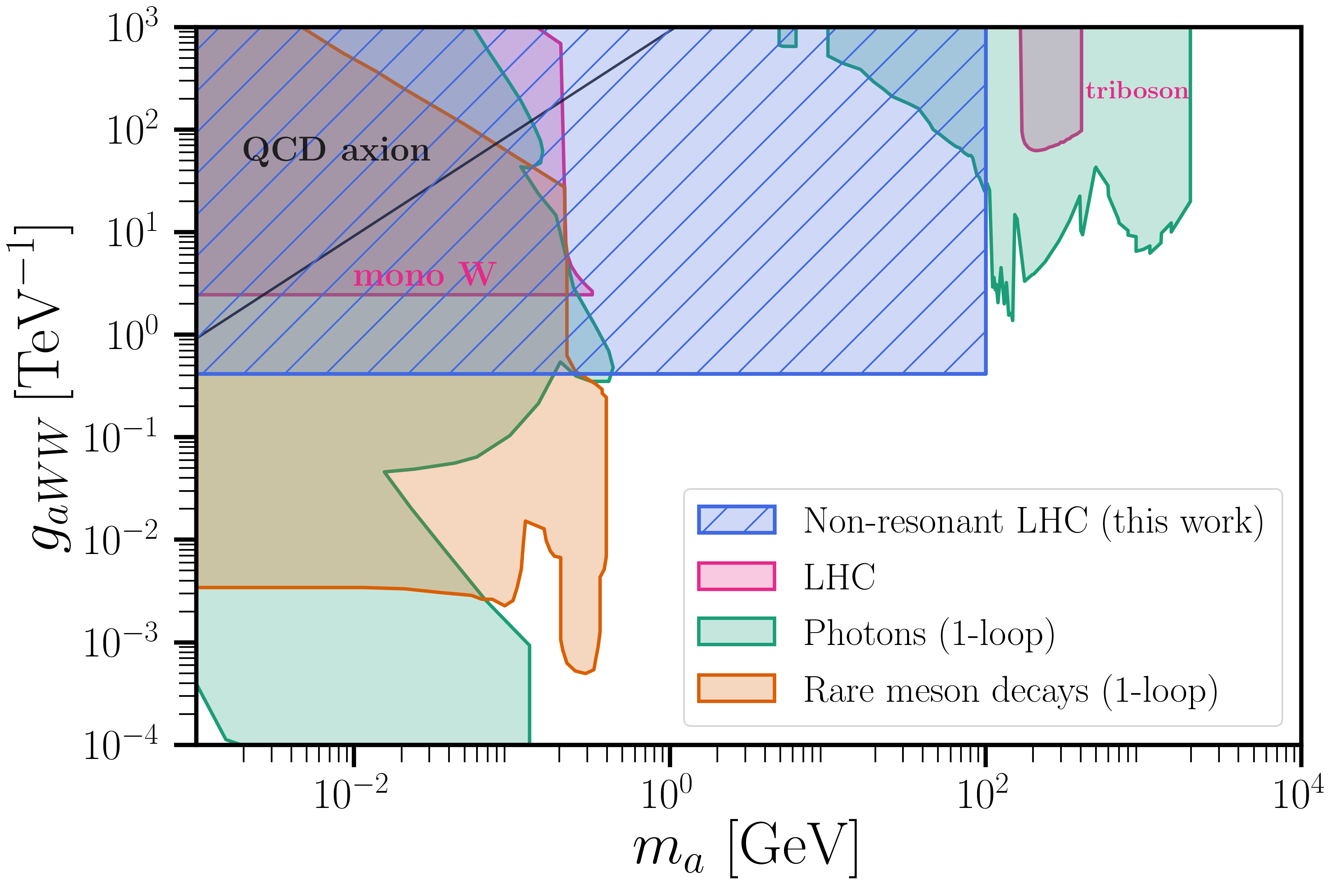}
    \includegraphics[width=\columnwidth]{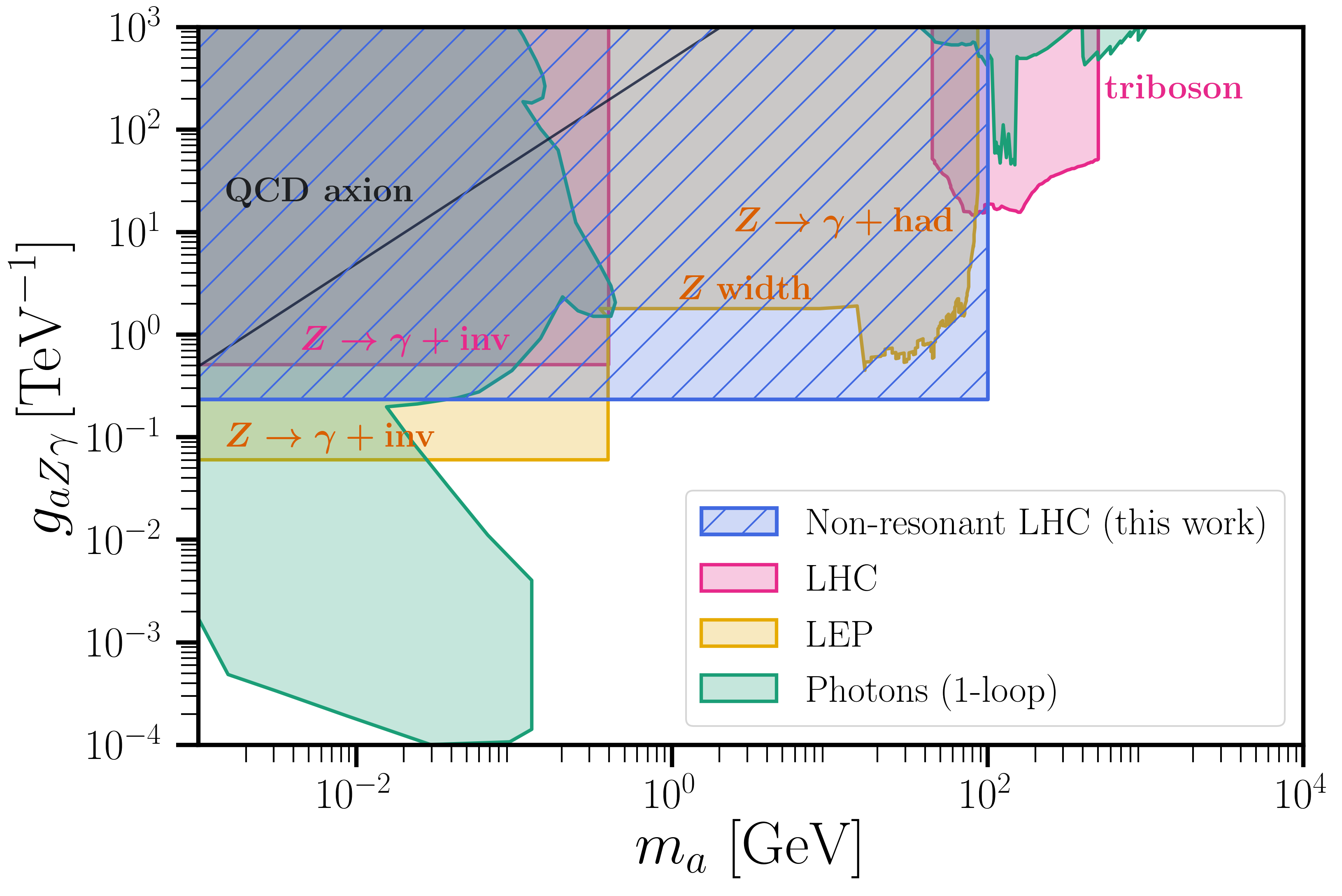}
    \caption{95\% CL constraints on $g_{aWW}$ (top) and $g_{aZ\gamma}$ (bottom). The constraints derived in this work are shown as the hatched area \PQs{with the assumption  $g_{aV_1V_2}/g_{agg}=\alpha_{V_1V_2}/\alpha_s$}. Also shown are constraints from other experiments, see text.
    }
    \label{fig:worldlimits}
\end{figure}

In Fig.~\ref{fig:worldlimits} the constraints derived in this analysis on $g_{aWW}$ and $g_{aZ\gamma}$ are compared to those from other experiments, see Ref.~\cite{Alonso-Alvarez:2018irt} and references therein. For this purpose, the assumption is made that $g_{agg}$ is related to the coupling to EW vector-bosons via the gauge coupling strengths, $g_{aV_1V_2}/g_{agg}=\alpha_{V_1V_2}/\alpha_s$ as discussed above.
The constraints labelled ``Photons" are based on beam dump experiments, supernova SN1987a observations, as well as LHC results. Due to radiative corrections of the axion-boson couplings to the axion-photon couplings, these results can be converted to constraints on $g_{aWW}$ and $g_{aZ\gamma}$, assuming only a very mild dependence on $f_a$~\cite{Alonso-Alvarez:2018irt}. For the constraints labelled ``LHC", it is a assumed that $g_{agg}$  is much larger than $g_{aWW}$ or $g_{aZ\gamma}$, eliminating much of the dependence on the gluon coupling~\cite{Alonso-Alvarez:2018irt}. 
The LHC bound on $g_{aZ\gamma}$ for low masses has been updated with respect to Ref.~\cite{Alonso-Alvarez:2018irt} with the recent results on $Z\rightarrow\gamma+\text{inv}$ by ATLAS \cite{ATLAS:2020uiq}. For reference purposes, the QCD axion line is shown in black assuming $\mathcal{O}(1)$ anomaly factors \footnote{i.e. $g_{a W W}=({\alpha_{e m}}/{s_{w}^{2}})/({2 \pi f_{a}}) $ and $g_{a Z \gamma}=({\alpha_{e m}}/{s_{w}c_w})/({2 \pi f_{a}}) $.}

For the $g_{aWW}$ coupling, the present analysis closes a gap in the coverage in the mass range between 0.4~GeV and 100~GeV which was a challenging region of masses since previous analyses studying solely EW axion couplings are no longer applicable in the presence of a gluonic coupling for masses above $m_a>3m_\pi$ due to the opening of the hadronic decay channel. For the $g_{aZ\gamma}$, a lot of the mass range covered by this analysis was already excluded by LEP analyses but this analysis extends to lower couplings by up to a an order of magnitude. 

We have evaluated recent ATLAS cross-section measurements of $WW$ and $Z\gamma$ in the search for off-shell production of low-mass axion-like particles. Such particles would alter the spectrum at high energies. In the absence of any significant excesses the measurements allow to constrain interactions of axion-like particles with EW and strong gauge bosons. Together with the previous constraints based on the analysis of $\gamma\gamma$ data from Ref.~\cite{Gavela:2019cmq}, the presented analysis provides strong constraints on the coupling parameters of the linear bosonic ALP EFT. The constraints presented here are the strongest to date in the mass range $m_a\sim (0.4-100)$~GeV for both the $aWW$ and the $aZ\gamma$ coupling for the parameter ranges considered.

\section*{Acknowledgements}
We thank B.~Gavela and V.~Sanz for the inspiration for this analysis, the useful discussions and for making the ALPs EFT UFO models available for use. We also thank Gonzalo Alonso-\'Alvarez for providing us the data of the combined previous bounds. The work by V.~Goumarre, B.~Heinemann, S.~Heim and P.~Qu\'ilez was in part funded by the Deutsche Forschungsgemeinschaft under Germany's Excellence Strategy - EXC 2121 ``Quantum Universe" - 390833306. S.~Heim thanks the Helmholtz Association for the support through the "Young Investigator Group" initiative. This work has benefited from computing services provided by the German National Analysis Facility (NAF).

\bibliographystyle{ieeetr}
\bibliography{biblio}

\end{document}